\documentclass[proof]{pasj00}
\SetRunningHead{Astronomical Society of Japan}{Usage of \texttt{pasj00.cls}}
\usepackage[authoryear,round,colon,compress]{natbib}
\usepackage{times}
\usepackage[dvips]{graphicx}
\draft

\begin{document}

\title{Suzaku Studies of the Supernova Remnant CTB~109 Hosting the Magnetar 1E~2259+586}

\author{
	\textsc{Toshio Nakano}$^1$,
	\textsc{Hiroaki Murakami}$^1$,
	\textsc{Kazuo Makishima}$^{1,2,3}$,
	\textsc{Junoko S. Hiraga}$^{3}$,
	\textsc{Hideki Uchiyama}$^4$,
	\textsc{Hidehiro Kaneda}$^5$,
	\textsc{and}
	\textsc{Teruaki Enoto}$^{6,7}$
	\thanks{Last update: October 23, 2014}
}

\affil{%
	$^1$Department of Physics, Graduate School of Science,The University of Tokyo 7-3-1 Hongo, Bunkyo-ku, Tokyo 113-0033\\
	$^2$MAXI Team, Institute of Physical and Chemical Research (RIKEN) 2-1 Hirosawa, Wako-shi, Saitama 351-0198\\
	$^3$Research Center for the Early Universe, The University of Tokyo 7-3-1, Hongo, Bunkyo-ku, Tokyo 113-0033\\
	$^4$Science Education, Faculty of Education, Shizuoka University 836 Ohya, Suruga-ku, Shizuoka 422-8529\\
	$^5$Goddard Space Flight Center, NASA, Greenbelt, Maryland, 20771, USA\\
	$^6$Graduate School of Science, Nagoya University, Chikusa-ku, Nagoya, 464-8602\\
	$^7$RIKEN Nishina Center, 2-1 Hirosawa, Wako-shi, Saitama 351-0198
}

\email{nakano@juno.phys.s.u-tokyo.ac.jp}

\maketitle
\begin{abstract}
	Ages of the magnetar 1E 2259+586 and
		the associated supernova	remnant CTB~109 were studied.
	 Analyzing the Suzaku data of CTB~109,
		its age was estimated to be $\sim$14~kyr,
		which is much shorter than the measured characteristic age of 1E 2259+586, 230 kyr. 
	This reconfirms the previously reported age discrepancy of this magnetar/remnant association,
		and suggests that the characteristic ages of magnetars are generally over-estimated as compared to their true ages. 
	This discrepancy is thought to arise because the former are calculated without considering decay of the magnetic fields. 
	This novel view is supported independently by much stronger Galactic-plane concentration of magnetars than other pulsars.
	The process of magnetic field decay in magnetars is mathematically modeled.
	It is implied that magnetars are much younger objects than previously considered,
		and can dominate new-born neutron stars.
\end{abstract}
\section{INTRODUCTION}
	\subsection{Relations between Magnetars and SNRs}
		\label{sec:MagSnr}
		Twenty seven Galactic and Magellanic X-ray sources are thought to constitute a class of objects called magnetars,
			which are single neutron stars (NSs) with extremely strong magnetic fields of $B=10^{14-15}$~G.
		They are believed to shine (mainly in X-rays) consuming the energies in their strong magnetic fields,
			because their X-ray luminosities exceed their spin-down luminosities 
			and they are not likely to be accreting objects.
		The magnetar concept well explains other peculiar characteristics of these objects, 
			such as long pulse periods clustered in a narrow range ($2-11$~s), 
			relatively large period derivatives, and unpredictable sporadic burst activities.
		However, we do not know yet how they are formed and how such strong fields evolved.
		
		Supernova remnants (SNRs) associated with magnetars are expected  to provide us 
			with valuable clues to the scenario of magnetar production (e.g., \citealt{Vink2008,SafiHarb2013}).
		As a result, the study of SNR/NS associations (e.g., \citealt{Seward1985,Chevalier2005, Chevalier2011}) 
			has been re-activated since the concept of magnetar has emerged.
		Although no clear difference  in the explosion energy has yet been found between SNRs with and without magnetars (e.g., \citealt{Vink2006}),
			an X-ray metallicity study of the SNR Kes~73,
			hosting the magnetar 1E~1841-045, led \citet{kumar2014}  to infer
			that the progenitor of this system had a mass of $\gtrsim 20~M_\odot$,	where $M_\odot$ is the solar mass.
		Through investigations of several SNR/magnetar associations, \citet{SafiHarb2013} characterized environments 
			that  are responsible for the magnetar production,
			and reinforced the view of rather massive progenitors.
		
		Apart from the progenitor issue,
			one particularly interesting aspect of magnetars,
			which can be studied by simultaneously considering the associated SNRs,
			is their age comparison.
		Of course, as discussed by \citet{Gaensler2004},
			the ages of a magnetar and of the associated SNR,
			estimated independently,
			must agree for them to be regarded as a true association.
		However,  these two age estimations sometimes disagree even in pairs with very good positional coincidence,
			including the 1E 2259+586/CTB~109 pair which is the topic of the present paper.
		This is often called ``age problem''.
		While \citet{Allen2004} suspected that the problem arises 
			because an SNR age estimate is affected by the presence of a magnetar,
			\citet{Colpi2000} instead attributed it to magnetic field decay of magnetars,
			which can make their characteristic ages much longer than their true ages.
		After the discovery of SGR 0418+5729 \citep{Rea2010a},
			a magnetar with a low dipole magnetic field, 
			the field's decay scenario has become more attractive
			\citep{DallOsso2012, Igoshev2012}.

		Through X-ray studies of CTB 109,
			the present paper attempts to address two issues.
			One is to solve the age problem with the 1E~2259+586/CTB~109 system,
			and the other is to conversely utilize the result to reinforce the nature of magnetars as magnetically driven NSs.
		After a brief introduction to the target system (section \ref{sec:1E2259CTB109}),
			we describe in  section~\ref{sec:Obs} and section~\ref{sec:DATAANA}
			 recent Suzaku observations of CTB~109, 
			and reconfirm the age problem in the system (section \ref{sec:AgeEstimation}).
		Then, an attempt is made in section~\ref{sec:SolveAge} to solve it invoking the decay of magnetic fields.
		Finally, we discuss some implications for the general view of NSs, 
			including in particular their magnetic evolution.
		Other topics will be discussed elsewhere, 
			including more detailed X-ray diagnostics of CTB 109,
			the origin of its peculiar half-moon shape, and characterization of the progenitor.	

\section{Magnetar 1E~2259+586 and SNR CTB~109}
	\label{sec:1E2259CTB109}
	\subsection{CTB~109}
		\label{sec:CTB109}
		The Galactic SNR, CTB~109,
			hosting the central point X-ray source 1E~2259+586,  
			was first discovered by the Einstein Observatory  \citep{Gregory1980} as an extended X-ray source with a peculiar semi-circular shape.
		It was independently identified as a shell-type SNR by radio observations at 610 MHz \citep{Hughes1981}. 
		A 10~GHz radio map taken with the Nobeyama Radio Observatory  revealed 
			good positional coincidence between the radio and X-ray shells, 
			while detected  no significant  enhancement from 1E~2259+586 \citep{Sofue1983}.
			
		Through CO molecular line observations, \cite{Heydari-Malayeri1981}, \cite{Tatematsu1985}, and \cite{Tatematsu1987}  found
			a giant molecular cloud located next to CTB~109,
			and suggested that it may have disturbed the SNR on the west side.
		\cite{Sasaki2004}  conducted a comprehensive X-ray study of this SNR with XMM-Newton.
		Assuming a distance of $D=3.0$~kpc \citep{Kothes2002,Kothes2012},
			 they estimated the shock velocity, age, and the explosion energy as $\upsilon_\mathrm{s} = 720 \pm 60$ km s$^{-1}$, 8.8 kyr, and $(7.4\pm 2.9)\times10^{50}$ erg respectively.
		Strong evidence for an interaction between the SNR shock front and the CO cloud was found by using $^{12}$CO, $^{13}$CO and Chandra observations \citep{Sasaki2006}.
			Furthermore, gamma-ray emission was detected with the Fermi-Lat  from CTB~109 \citep{Castro2012}.
		Finally, using Chandra, \cite{Sasaki2013} detected emission from the ejecta component and refined the age as 14~kyr.
		
	\subsection{1E~2259+586}
		\label{sec:1E2259}
		The compact object 1E~2259+586 was first detected in X-rays nearly at the center of CTB~109 \citep{Gregory1980}.
		It was soon found to be a pulsar, and the pulse period was at first considered as $P=3.49$~s \citep{Fahlman1982}.
		This was due to the double-peaked pulse profile, and the fundamental period was soon revised to $P=6.98$~s \citep{Fahlman1983}.
		Repeated X-ray observations enabled the spin down rate to be measured as $\dot{P}=(3-6)\times 10^{-13}~\mathrm{ ss^{-1}}$ \citep{Koyama1987,Hanson1988,Iwasawa1992}, and 
			these results made it clear that the spin-down luminosity of 1E~2259+586  ($5.6 \times 10^{31}$~erg~s$^{-1}$) is far insufficient to explain its X-ray luminosity, $1.7\times 10^{34}$~erg~s$^{-1}$.
		Due to this and the long pulse period, 
			1E~2259+586 was long thought to be an X-ray binary with an orbital period of $\sim 2300$~s  (e.g. \citealt{Fahlman1982}),  and
			extensive search for a counterpart continued \citep{Davies1989,Coe1992,Coe1994}.
		However,  no counterpart was found \citep{Hulleman2000},
			and instead, tight upper limits on the orbital Doppler-modulation have been obtained 
			as $a_x \sin i < 0.8$ light-s \citep{Koyama1989},  $a_x \sin i < 0.6$ light-s \citep{Mereghetti1998}, and $a_x \sin i < 0.028$ light-s \citep{Baykal1998}.
		Here, $a_x$ is the semi-major axis of the pulsar's orbit, and $i$ is the orbital inclination.
		These strange properties of 1E~2259+586 led this and a few other similar objects to be called Anomalous X-ray Pulsars (AXPs).
	
		In the 1990's, several attempts were made to explain 1E~2259+586 without invoking a companion:
			e.g., massive white dwarf model \citep{Usov1994}, and precessing white dwarf model \citep{Pandey1996}. 
		Monthly observations of 1E~2259+596 over 2.6 years with RXTE gave phase-coherent timing solutions indicating a strong stability over that period \citep{Kaspi1999}.
		This favored non-accretion interpretation.
		\cite{Heyl1999}  suggested that spin-down irregularities of AXPs are statistically similar to glitches of radio pulsars.
		Meanwhile, the concept of magnetar was proposed to explain Soft Gamma Repeaters (SGRs) as magnetically-powered NSs, namely, magnetars \citep{Duncan1992,Thompson1995}.
		Furthermore, like SGRs, 1E~2259+586 showed an X-ray outburst \citep{Kaspi2002, Gavriil2004, Woods2004}.
		Today, AXPs including 1E~2259+586, as well as SGRs, are both considered to be magnetars. 		
		Employing the canonical assumption of spin down due to magnetic dipole radiation, the measured pulse period of $P=6.98$~s \citep{Iwasawa1992},
			 and its derivative, $\dot{P}=4.8\times10^{-13}$ss$^{-1}$ (e.g., \citealt{Gavrill2002}), give a dipole magnetic field of  $5.9\times 10^{13}$~G and a characteristic age of  $\tau_\mathrm{c} = P/2\dot{P} = $230~kyr.
	
\section{OBSERVATIONS AND DATA REDUCTION}
	\label{sec:Obs}
	The Suzaku observations of 1E~2259+586 and CTB~109 were made on two occasions. 
	The first of them was conducted on 2009 May 25 as a part of the AO4 Key Project on magnetars \citep{Enoto2010b}.
	The three cameras (XIS0, XIS1, and XIS3) of the X-ray Imaging Spectrometer (XIS) onboard Suzaku were operated in 1/4-window mode with 
		a time resolution of 2~s \citep{koyama2007}, to study the 6.98~s  pulsation.
	The rectangular $(17' \times 4'.3)$ fields of view of the 1/4-window were center on this magnetar,
		while the SNR was partially covered.
	As the second observation, 
		we performed new four pointings onto CTB~109 with Suzaku (PI:T.Nakano) on 2011 December 13 .
	In order to synthesize a whole image of the SNR, all XIS cameras were operated in full-window mode under the sacrifice of time resolution.
	A log of these observations is given in Table \ref{tb:ObservationTable}.
	
	In the present work, we use the XIS data which were prepared with version 2.7.16.31 pipeline proceeding, and the calibration data updated in January 2013. 
	The data of the Hard X-ray Detector (HXD) from neither observation are utilized here, 
		since the Suzaku data analysis in the present paper focuses on CTB~109,
		rather than the central magnetar.
	The data reduction  was carried out  using the HEADAS software package version 6.13, and spectral fitting was performed with {\tt xspec} version 12.8.0.
	The redistribution matrix files and the auxiliary response files of the XIS were generated  with  {\tt xisrmfgen}  and {\tt xissimarfgen} \citep{Ishisaki2007} respectively. 

	\begin{table}[htbp]
		\caption{ {\normalsize Log of Suzaku observations of CTB~109.}}
			\centering
			\begin{tabular}{ccccc}
				\hline \hline
				Observation ID & $\alpha$ & $\delta$ & Start Time & Exposure( ks )\\ \hline
				404076010 & 23$^\mathrm{h}$ 01$^\mathrm{m}$ 04$^\mathrm{s}$.08 & $58^\circ 58'15''.6$ & 2009-05-25 20:00:17 & 122.6 \\
				506037010& 23$^\mathrm{h}$ 01$^\mathrm{m}$ 06$^\mathrm{s}$.96	& $59^\circ 00 ' 50''.4$& 2011-12-13 06:48:41 & 40.8 \\
				506038010 & 23$^\mathrm{h}$ 00$^\mathrm{m}$ 26$^\mathrm{s}$.88	& $58^\circ44' 13''.2$ & 2011-12-14 04:47:02 & 41.4 \\
				506039010 & 23$^\mathrm{h}$ 03$^\mathrm{m}$ 06$^\mathrm{s}$.96	& $58^\circ58'51''.6$ & 2011-12-15 01:57:25 & 30.4 \\									
				506040010 &23$^\mathrm{h}$ 03$^\mathrm{m}$ 06$^\mathrm{s}$.96 & $58^\circ 40' 51''.6$ & 2011-12-15 18:03:52 & 30.5\\
				\hline
			\end{tabular}
		\label{tb:ObservationTable}
	\end{table}%

\section{DATA ANALYSIS AND RESULT}
\label{sec:DATAANA}
\subsection{Image Analysis}
	\label{sub:ImageAna}
\begin{figure}[htbp] 
		\centering
		\includegraphics[width=5in]{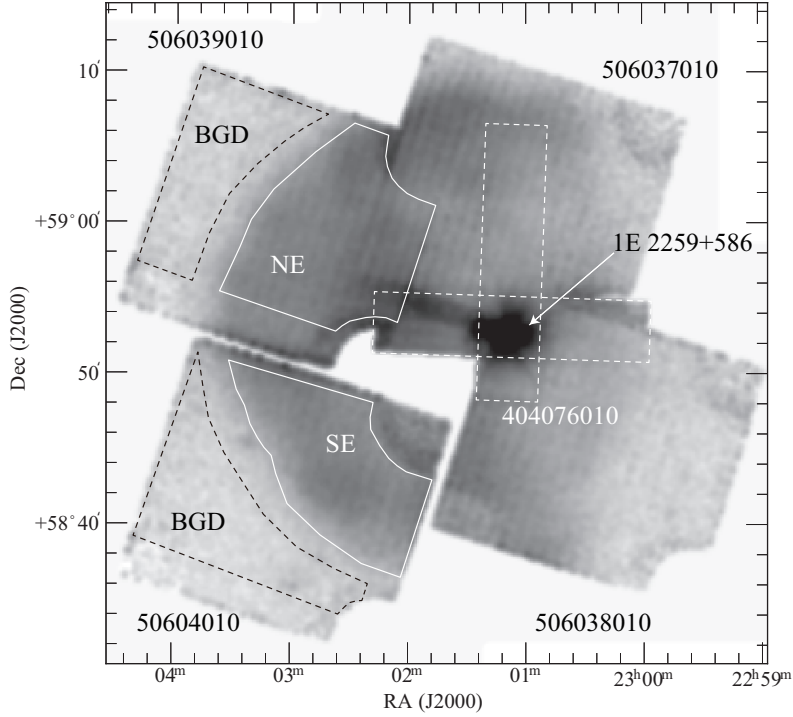} 
	  	\caption{
			A mosaic image of CTB~109 obtained in 0.4--5.0~keV with the Suzaku XIS.
			After subtracting the non X-ray background, the image was corrected for exposure and vignetting.
			The on-source and background regions are indicated by white and dashed black lines, respectively.
			Some corners of the square XIS fields of view are masked to remove calibration isotopes.
		}
	  	\label{fig:XisImage}
	\end{figure}

	Figure \ref{fig:XisImage} shows a gray-scale mosaic image of CTB~109 obtained with the Suzaku XIS,
		shown after subtracting non X-ray background.
	The magnetar, 1E~2259+586, appears as a bright source at the center.
	The crisscross region  including 1E~2259+586 was taken in the first observation, 
		while four square regions represent those from the second one.
	Thus, the mosaic XIS image reconfirms the half-moon like morphology of this SNR (section\ref{sec:CTB109}).

	As mentioned in section \ref{sec:CTB109}, the lack of western part of CTB 109 is usually attributed to its interactions with giant molecular clouds.
	Bright regions in the SNR are also thought to be the signature of such an interaction \citep{Sasaki2006}.
	These issues, together with detailed spatial distributions of X-ray spectral properties,  will be postponed to another publication. 
	The present paper instead deals with average X-ray spectra, 
		because our prime motivation is to estimate the age of CTB~109.

\subsection{Spectral Analysis}
\label{subsec:Spect}
\begin{figure}[htbp] 
		 \centering
		 \includegraphics[width=6.5in]{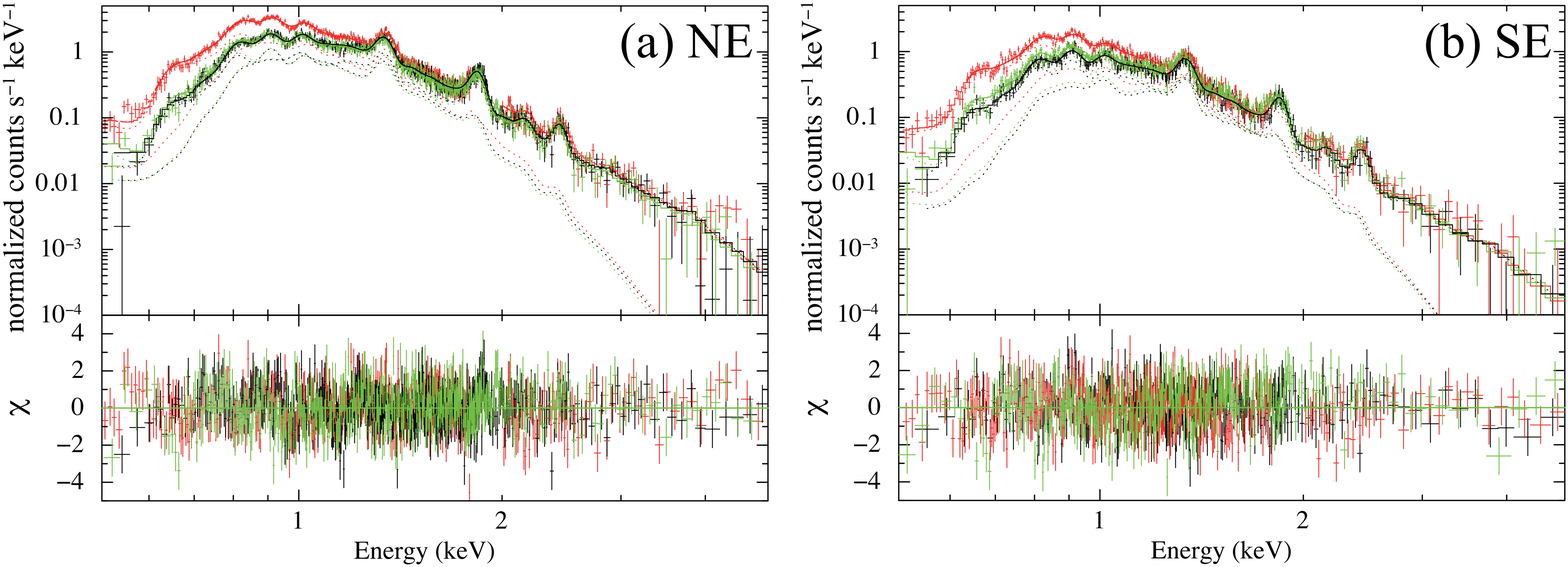} 
		 \caption{Background-subtracted XIS0~(black), XIS1~(red) and XIS3~(green) spectra of CTB~109, extracted from the NE (panel a) and SE (panel b) regions.
		 	They are fitted simultaneously with a two-component model described in the text. 
			The dotted lines indicate individual model components,
				with the three colors specifying the three cameras.		
		}
	\label{fig:Spect}
	\end{figure}

	Figure \ref{fig:Spect} shows XIS0, XIS1 and XIS3 spectra of CTB~109 from the second observation,  
		extracted from the regions indicated in figure \ref{fig:XisImage}   as ``NE'' and ``SE'', which represent two eastern pointing positions. 
	Since these eastern parts of the SNR has kept smooth round shape, the effect from the interaction with GMC seems to be small. 
	Background spectra were extracted from source free regions in the same observation which are also indicated as ``BGD'' in figure \ref{fig:XisImage}.
	Spectral bins were summed up to attain a minimum of 30 counts bin$^{-1}$. 
	The "NE" and "SE" spectra are very similar and both exhibit emission lines due to highly ionized atoms such as 
		Ne-IX triplet ($\sim$0.92~keV),  Ne-X Ly$\alpha$ (1.02~keV), 
		Mg-XI triplet ($\sim$1.35~keV), Si-XIII triplet ($\sim$1.87~keV), and S-XV triplet ($\sim$2.45~keV).
	
	We first applied  a variable-abundance non-equilibrium ionization~({\tt VNEI}) plasma emission model to the NE spectra in figure \ref{fig:Spect} (a).
	However, even when abundances of Ne, Mg, Si and S are allowed to vary freely,   the reduced $\chi^2$ of the fit was not made lower than 1.5.
	Thus, the  single temperature VNEI model was rejected.
	Other plasma models in {\tt xspec}, such as  {\tt vpshock} and {\tt vmekal} were also unsuccessful.

	Then, we considered that the SNR emission consists of two components including ejecta and shocked inter-stellar medium (ISM), 
		and added a non-equilibrium ionization (NEI) plasma emission model to model shocked ISM component.
		Abundances of ISM component was fixed to solar  to reduce the number of free parameters. 

	
	By introducing this two-component emission model, the fit was improved to  $\chi^2/\nu = $1.28. 
	Since different XIS cameras gave discrepant fit residuals around Mg-XI K$\alpha$ and Si-XIII K$\alpha$ lines, 
		presumably due to calibration uncertainties of the XIS,
		we allowed gain parameters of the XIS cameras to vary. 
	Then, the fit has become acceptable with $\chi^2/\nu$ = 1.06 (1293/1223).
	The obtained best-fit parameters are shown in table \ref{tab:parameters},
		while the gain parameters in table \ref{tab:gainparameters}.
	Since the obtained abundances of the second (Plasma~2) component are all consistent with 1 solar, it is also likely to be dominated by the ISM.
	The two components are both inferred to be somewhat deviated from ionization equilibrium.
	The fit was not improved even when the abundance of the Plasma~2 component is allowed to change from 1.0.
	Similarly, we analyzed the SE spectra shown in figure \ref{fig:Spect} (b), 
		and derived the parameters which are again summarized in table \ref{tab:parameters}.
	Thus, the two temperatures are in good agreement between the two regions, 
		although the SE region gives somewhat higher metal abundances.
	This two-component emission model was also employed by \cite{Sasaki2013}. 
	\begin{table}[htb]
		\begin{center}
			\caption{Best-fit spectrum parameters for the NE and SE regions\footnotemark[$*$].}
			\begin{tabular}{lcrrc} \hline \hline
			Component & Parameter & NE     & SE \\
			\hline
			Absorption & $N_\mathrm{H} \left(\mathrm{10^{22}~cm^{-2}}\right)$ &  $0.78\pm0.01$  &  $0.73^{+0.01}_{-0.02}$  \\
			Plasma 1 (VNEI) & $kT_\mathrm{1}$ &  $0.62^{+0.04} _{-0.01}$ &  $0.65^{+0.02}_{-0.01}$  \\
			&  $\eta_1$\footnotemark[$*$] $ \left( 10^{12}~\mathrm{cm^{-3}~s}\right)  $  &  $0.37^{+0.05}_{-0.04}$  & $0.22^{+0.05}_{-0.06}$  \\
			&  $\mathrm{Ne}$ (solar)  &  $0.8\pm0.2$  &  $0.8^{+0.2}_{-0.1}$  \\ 
			&  $\mathrm{Mg}$ (solar)  &  $0.88^{+0.01}_{-0.06}$  &  $1.1\pm{0.1}$  \\ 
			&  $\mathrm{Si}$ (solar)  &  $1.2\pm{0.1}$  &  $1.7^{+0.1}_{-0.3}$  \\ 
			&  $\mathrm{S}$ (solar)  &  $1.0\pm{0.1}$  & $1.7^{+0.2}_{-0.3}$  \\ 
			&  $\mathrm{Fe}$ (solar)  &  $0.99^{+0.05}_{-0.06}$  &  $1.0^{+0.2}_{-0.1}$  \\
			&  $K_1$\footnotemark[$\dagger$] ($10^{-2}~\mathrm{cm}^{-5}$)  	& $3.7^{+0.2}_{-0.1}$   & $1.2\pm0.1 $  \\
			Plasma 2 (NEI)& $kT_\mathrm{2}$ &  $0.26\pm0.01$  & $0.25\pm0.01 $ \\
			&  $\eta_2$\footnotemark[$*$]  $\left( 10^{12}~\mathrm{cm^{-3}~s}\right)  $  &  $>1$ &  $>1$ \\
			&  Abundance (solar)  &  1 (fixed)  &1 (fixed)  \\
			&  $K_2$\footnotemark[$\dagger$] ($10^{-2}~\mathrm{cm}^{-5}$)   &  $19.6^{+0.8}_{-1.4}$  & $13.2^{+0.8}_{-1.6}$  \\
			\hline
			$\chi^2$/dof  &  &  1294/1223  &  1138/951  \\
			\hline 
				\multicolumn{4}{@{}l@{}}{\hbox to 0pt{\parbox{170mm}{
				\par\noindent
				\footnotemark[$*$]Uncertainties are statistical errors at 90 \% confidence.
				\\
				\footnotemark[$*$] Ionization parameter, defined as  $\eta = n_\mathrm{e} t$ \\
				\footnotemark[$\dagger$] Normalization of NEI or VNEI, defined as $K=\frac{10^{-14}}{4\pi D^2} \int n_\mathrm{e} n_\mathrm{H} dV $
			}\hss}}
 			\end{tabular}
			\label{tab:parameters}
		\end{center}
	\end{table}

	\begin{table}[htb]
		\begin{center}
			\caption{The best-fit gain parameters\footnotemark[$*$].}
			\begin{tabular}{llcc}\hline\hline
				region & instrument & slope & offset (eV)\\ \hline
				NE & XIS0 & 0.997 $\pm 0.001$ & $-5.3 \pm 0.2  $  \\
				& XIS1 & 1.004 $\pm0.001$ & $  -3.0 \pm 0.3 $  \\
				& XIS2 & 0.986 $\pm0.001$&  $  -7.0 \pm 0.4$ \\ \hline
				SE & XIS0 & 1.003 $\pm 0.001$ & $-10.6 \pm 0.1 $  \\
				& XIS1 & 0.993 $\pm0.001$ & $  6.6 \pm 0.1  $  \\
				& XIS2 & 0.985 $\pm0.001$&  $  6.6 \pm 0.2 $  \\ \hline

				\multicolumn{4}{@{}l@{}}{\hbox to 0pt{\parbox{170mm}{
				\par\noindent
				\footnotemark[$*$]Uncertainties are statistical errors at 90 \% confidence.
			}\hss}}

			\end{tabular}
			\label{tab:gainparameters}
		\end{center}
	\end{table}
	
\section{Estimation of the Age of CTB~109}
\label{sec:AgeEstimation}
	Now that the average plasma properties of CTB~109 have been quantified, 
		let us proceed to our prime goal of studying this SNR, 
			i.e., its age estimation.
	First, the physical radius of CTB~109 is estimated as $R = (16 \pm 1 )d_{3.2}$~ pc , from its angular size of $\sim 16'$
		and the estimated distance $D=3.2\pm0.2$~kpc \citep{Kothes2012}. 
	Here, $d_{3.2} = D/3.2$ is scale factor of distance.
	Next, CTB~109 may be considered in the Sedov phase (neglecting the missing western part).
	Then, applying the Sedov-Taylor similarity solution \citep{Sedov1959,Taylor1950}  to this SNR,
		its age can be obtained as
	 \begin{equation}
		\tau_\mathrm{SNR} = \frac{2}{5} \frac{R}{\upsilon_\mathrm{s} }
		\label{eq:SedovAge}
	 \end{equation} 
	where $\upsilon_\mathrm{s}$ again represents the shock front velocity.
	
	Assuming the strong shock limit, we can calculate $\upsilon_\mathrm{s}$ from the post-shock temperature $T_\mathrm{ps}$ as
	\begin{equation}
		\upsilon_\mathrm{s} = \sqrt{ \frac{16}{3\bar{m}} kT_\mathrm{ps}}
	\end{equation}
	where $k$ and $\bar{m}$, respectively  represent the Boltzmann constant and the mean mass per free particle.
	Assuming a solar abundance (section~\ref{subsec:Spect}), we employed $\bar{m} \simeq 0.61~m_\mathrm{p}$
	where $m_\mathrm{p}$ is the proton mass.
	Since CTB~109 is a middle-aged SNR without too strong shocks,
		the electron temperature of the NEI component measured  in section \ref{sec:DATAANA} can be considered to be close to the kinematic ISM temperature,
		and hence to $T_\mathrm{ps}$ \citep{Ghavamian2007}. 
	Thus, substituting $0.25 \pm 0.02$~keV for $T_\mathrm{ps}$,
		 equation (2) gives $\upsilon_\mathrm{s}=460\pm40$~km~s$^{-1}$, 
		 and then equation (1) yields $\tau_\mathrm{SNR}\simeq(14\pm2) d_{3.2}$~kyr in agreement with \cite{Sasaki2013}.
	Compared to the estimated $\tau_\mathrm{SNR}$, 
		the characteristic age of 1E~2259+586, $\tau_\mathrm{c} = 230$~kyr (section~\ref{sec:1E2259}),
		is $\sim$16 times larger.

	If we assume that CTB~109 is in cooling phase rather than Sedov phase,
		the time dependence of the radius becomes $R \propto t^{2/7}$ \citep{McKee1977}.
	Then, the age estimation slightly changes to
	\begin{equation}
		\tau_\mathrm{SNR} = \frac{2}{7} \frac{R}{\upsilon_\mathrm{s} },
		\label{eq:Cooling}	
	\end{equation} 
	and $\tau_\mathrm{SNR} = (10\pm1) d_{3.2}$~kyr is obtained.
	Therefore the age discrepancy still persists (even increases).
		
		Let us cross-check the above estimates using the parameter $\eta\equiv n_\mathrm{e}t$ (table \ref{tab:parameters}) of the ejecta component. 	
		 We assume that the SNR has a half spherical shell with thickness of $\Delta R = R/12$ (assuming $4/3 \pi R^3 n_0 m_\mathrm{p} =  4\pi R^2 \Delta R n_0 m_\mathrm{p}$ ), 
		 	and the ISM component corresponding to Plasma~2 in table \ref{tab:parameters} is emitted from this shell.
		Furthermore, as a simplest approximation,  the ejecta component (Plasma~1 in table \ref{tab:parameters}) may be assumed to uniformly fill the inner region,
			considering that the reverse shock has reached the center of the SNR.
		In the NE region, the extracted emission volumes 				
			of Plasma~1(ejecta) and Plasma~2(ISM) can be obtained numerically as $V_1 = 4.0\times 10^{58}d_{3.2}^3$~cm$^{3}$ and $V_2 = 1.4\times 10^{58} d_{3.2}^3$~cm$^{3}$, respectively.
		Then	, the spectrum normalizations in table \ref{tab:parameters} give 
			the averaged density of the ejecta as $n_1 = (0.33\pm0.03) d_{3.2}^{-1/2}$~cm$^{-3}$ and that of the ISM-shell as $n_2=(1.2\pm0.1) d_{3.2}^{-1/2}$~cm$^{-3}$.
		The time required for the ejecta to become ionized as we now observe is hence estimated as 
		$\eta/n_1 = (29-34) d_{3.2}^{1/2}$~kyr.
		Applying the same argument to the SE region having the emission volumes of $V_1=6.7\times 10^{57} d_{3.2}^{3}$~cm$^{3}$ and $V_2 = 3.3\times 10^{57} d_{3.2}^{3}$~cm$^{3}$,   
			we obtain $n_1 = (0.46\pm0.05) d_{3.2}^{-1/2}$~cm$^{-3}$  and $n_2 = (0.55\pm0.06) d_{3.2}^{-1/2} $~cm$^{-3}$, and $\eta/n_1 = (11-19) d_{3.2}^{1/2} $~kyr.
		Pre-shock density is estimated as $n_0 = n_2/4 =  (0.1-0.3) d_{3.2}^{-1/2}$~cm$^{-3}$.
		Even though these estimates of $\tau_\mathrm{SNR}$ must have a certain range of systematic uncertainties, 
			large discrepancies significantly remain between $\tau_\mathrm{SNR}$ and $\tau_c$.
		In fact, it would be difficult to think 
			that CTB~109 are emitting X-rays even at an age of ~230~kyr while keeping the regular shape (except the missing western half),
			because its density environment as estimated is quite typical of a solar neighborhood.
		Thus, we reconfirm the previously reported age discrepancy \citep{Sasaki2013} between CTB~109 and 1E~2259+586.		

\section{Solving the Age Discrepancy}
\label{sec:SolveAge}
	The magnetar 1E~2259+586 is located nearly at the very center of the half-moon-shaped shell of CTB~109 (section~\ref{sub:ImageAna}, figure \ref{fig:XisImage}).
	This coincidence is difficult to explain by invoking a chance superposition of the two objects.
	We therefore assume that 1E~2259+586 and CTB~109 were indeed produced by the same supernova explosion (section \ref{sec:MagSnr}),
		while $\tau_\mathrm{c}$ of 1E~2259+586  is somehow significantly overestimated,
		compared to its true age which we consider to be close to $\tau_\mathrm{SNR}$.

\subsection{Case with a  Constant Magnetic Field}
	\label{subsec:constB}

	To solve the issue of the suggested overestimate of $\tau_\mathrm{c}$ 
		after \cite{Colpi2000} and \cite{DallOsso2012}, let us begin with reviewing the meaning of $\tau_\mathrm{c}$.
	In general, the spin evolution of a pulsar with dipole surface magnetic field $B$ is expressed empirically  as
	\begin{equation}
		\frac{d\omega}{dt} =  - b  B^2 \omega^{n}
		\label{eq:domega/dt}
	\end{equation}
	with $b \equiv 32\pi^3R_\mathrm{psr}^6/3I\mu_0 c^3$ and the braking index of $n=3$,  
		where $R_\mathrm{psr}=10$ km is the pulsar's radius,  
		$I=9.5\times 10^{44}~\mathrm{g~cm^2}$ its momentum of inertia,  $\mu_0$ vacuum permeability and $c$ the light velocity.
	If we use the pulse period $P=2\pi/\omega$ and its derivative $\dot{P}$ instead of $\omega$ and $\dot{\omega}$, 
		equation (\ref{eq:domega/dt}) becomes 
	\begin{equation}
		B = \sqrt{ \frac{P\dot{P}}{b} } \simeq 3.2 \times 10^{19} \sqrt{P\dot{P}} \ \mathrm{G}.
	\end{equation}
	If $B$ does not depend on time $t$,
		equation (\ref{eq:domega/dt}) can be integrated as
	\begin{equation}
		t  = -\frac{1}{n-1}\left( \frac{\omega}{\dot{\omega}} \right)\left[ 1 -\left( \frac{\omega}{\omega_0} \right)^{n-1}\right] 
		= \tau_\mathrm{c}\left[  1 - \left( \frac{\omega}{\omega_{0}} \right)^{n-1} \right]
		\label{eq:eqB1}
	\end{equation}	
	where $\omega$ and $\dot{\omega}$ both refer to the present values, while $\omega_{0}$ is the angular frequency at $t=0$ (i.e. the birth).
	Assuming that  	$\left( \omega/ \omega_0\right)^{n-1}$ can be neglected, 
		the characteristic age is defined as
	\begin{equation}
		\tau_\mathrm{c} \equiv  \frac{\dot{\omega}}{\left( n-1 \right) \omega}  \equiv \frac{P}{ \left( n-1 \right)  \dot{P}}.
		\label{eq:Taucdef}
	\end{equation}
	These equations are generally used for pulsars, and found in some textbooks (e.g., \citealt{PulsarAstro1998}).
	
	More generally,  the true age of the pulsar, denoted by $t_0$, can be compared with its $\tau_\mathrm{c}$ as
	\begin{equation}
		\frac{\tau_\mathrm{c}}{t_0} = \frac{1}{1-\left( \frac{\omega}{\omega_0}\right)^{n-1} } = \frac{1}{1-\left( \frac{P_0}{P}\right)^{n-1} } \simeq 1 + \left( \frac{P_0}{P} \right)^{n-1} ,
		\label{eq:eqB2}
	\end{equation}
	where $P_0 = 2\pi/\omega_0$,
		and the last expression is the first-order approximation in $\left( P_0/P\right)^{n-1}$.
	Thus, $\tau_\mathrm{c}$ becomes somewhat larger than $t_0$ if $\left(P_0/P\right)^{n-1}$ cannot be neglected.
	Conversely, if we somehow have an independent estimate of $t_0$ its comparison with $\tau_\mathrm{c}$ can be used to infer $P_0$ as
	\begin{equation}
		P_0 = P \left( - \frac{t_0}{\tau_\mathrm{c}} +1 \right)^{1/(n-1)}.
		\label{eq:eqB3}
	\end{equation}
	For example, the Crab pulsar \citep{Staelin1968}, with $P=33$ ms, $\dot{P}=2.42\times 10^{-13}~\mathrm{ss^{-1}}$ and $n$ = 2.509 \citep{Lyne1993},
		has $\tau_\mathrm{c} = 1241$~yr.
	Comparing this with its true age of ${\rm 960~yr}$ (as of 2014), 
		equation (\ref{eq:eqB3}) yields $P_0 = 18~{\rm ms}$ if assuming $n=2.509$, 
		on $P_0 = 15.7~{\rm ms}$ if  $n=3.0$ (for ideal magnetic dipole radiation). 
	Thus, regardless of the employed value of $n$, 
		the small difference between $\tau_\mathrm{c}$ and $t_0$ of the Crab pulsar can be understood to imply 
		that it has so far lost $\sim3/4$ of its initial rotational energy in $\sim 1~{\rm kyr}$.
	
	In contrast to the above case of young active pulsars,
		we would need to invoke $P_0 = P \times 0.97 = 6.76$ s,
	if equation (9) with $n=3$ were used to explain the large discrepancy,
	$\tau_\mathrm{c}/t_0 \sim \tau_\mathrm{c}/\tau_\mathrm{SNR}\sim 16$,
	found in the CTB/1E2259+586 system.
	This would lead to a view that 1E~2259+586 was born some 14~kyr ago as a slow rotator 
	of which the spin period is much longer than those (0.2~s to 2~s) of
	the majority of {\it currently} observed (hence relatively old) radio pulsars,
		and has so far lost only a tiny fraction of its rotational energy in 14~kyr.
	However, such a view is opposite to a general consensus
	that new-born magnetars must be rotating rapidly,
	even faster  than ordinary pulsars,
	in order for them to acquire the strong magnetic fields
	(e.g., \citealt{Usov1992},  \citealt{Duncan1996}, \citealt{Lyons2010}).
	Furthermore, an NS with $P_0 =6.67$ s,
		has an angular momentum of only $\sim 10^{-5}$ of those of typical new-born pulsars with $P_0 \sim 10$ ms including the Crab pulsar,
		and hence would require an extreme fine tuning 
		in the progenitor-to-NS angular momentum transfer during the explosion.
	We therefore conclude that the age problem of 1E2259+586
	cannot be solved as long as its magnetic field is assumed
	to have been constant since its birth.

\subsection{Effects of  Magnetic Field Decay}
\label{subsec:decay}
	Since the age problem of 1E~2259+586/CTB~109 cannot be solved as long as $B$ is considered constant,
		we may next examine the case where $B$ decays with time (section \ref{sec:MagSnr}).
	In fact, the X-ray emission of magnetar is thought to arise when their magnetic energies are consumed \citep{Thompson1995}.
	Then, the calculations presented in section 5.1 would be no longer valid, 
		and we need to integrate equation (\ref{eq:eqB1}) considering the time evolution of $B$.

	Let us consider a simple magnetic filed decay model employed by \cite{Colpi2000},
		namely
	\begin{equation}
		\frac{dB}{dt} = - a B^{1+\alpha}, 
		\label{eq:dB/dt}
	\end{equation}
	where  $\alpha\geq0$ is a parameter, and $a$ is another positive constant.
	This equation can be solved as 
	\begin{eqnarray}
		B\left(t\right) = \left\{ \begin{array}{ll}
		{\displaystyle \frac{B_0}{ \left( 1+ \frac{ \displaystyle  \alpha t}{ \displaystyle  \tau_\mathrm{d}} \right)^{1/\alpha} }}   &  \left( \alpha \neq 0 \right) \\  
		&	\\
		{\displaystyle B_0 \exp{ \left(  \frac{-t}{\tau_\mathrm{d}}\right) }}  & \left( \alpha = 0 \right)
		\end{array} \right.
		\label{eq:B(t)}
	\end{eqnarray}
	where  $B_0$ represents the initial value of $B$,  and $\tau_\mathrm{d} = \left(1/a B^\alpha_0\right)$, 
		an arbitrary constant,
		 means a typical lead time till the power-law like decay of $B$ begins.

	Substituting equation (\ref{eq:B(t)}) into equation (\ref{eq:domega/dt}), 
		we can derive $P$ as a function of $t$.
	Then, as already given by \cite{DallOsso2012}, 
		$\tau_\mathrm{c}$ can be expressed as a function of $t$, $P_0$ , $\alpha$ and $\tau_\mathrm{d}$ as 
	\begin{eqnarray}
		\tau_\mathrm{c} = \left\{ \begin{array}{ll}
		{\displaystyle \frac{\tau_\mathrm{d}}{2-\alpha}  \left[ \left\{  1 + \left(2-\alpha \right) \frac{\tau_0}{\tau_\mathrm{d}}   \right\}
			 \left(  1+\frac{\alpha t}{\tau_\mathrm{d}}\right)^{2/\alpha} - \left(1+\frac{\alpha t}{\tau_\mathrm{d}} \right) \right]}    & \left( \alpha \neq 0,2 \right)  \\
		&	\\ 
		{\displaystyle \frac{\tau_\mathrm{d}}{2}  \left[  \left( 1+ \frac{2\tau_0}{\tau_{\tau_\mathrm{d}} } \right) 
			 \exp{\left( \frac{2t}{\tau_\mathrm{d}}\right)} -1  \right] }   & \left( \alpha = 0  \right)   \\
		&	\\
		{\displaystyle  \left( 1+\frac{2t}{\tau_\mathrm{d}}\right) \left[\tau_0 + \frac{\tau_\mathrm{d}}{2} \ln{\left(1+\frac{2t}{\tau_\mathrm{d}} \right)} \right] }   
		& \left( \alpha = 2 \right)\\
		\end{array} \right.
		\label{eq:Tauc}
	\end{eqnarray}	
	where $\tau_0 \equiv P_0/2\dot{P}_0$ is the initial value of $\tau_\mathrm{c}$.
	The first form of equation (\ref{eq:Tauc}) reduces to equation (\ref{eq:eqB2}) for  $\alpha \rightarrow \infty$ or $\tau_\mathrm{d} \rightarrow \infty$, 
		i.e., the case of a constant $B$.

\subsection{Magnetic Field Evolution of 1E~2259+586}
\begin{figure}[htbp] 
		\centering
	   		\includegraphics[width=6.5in]{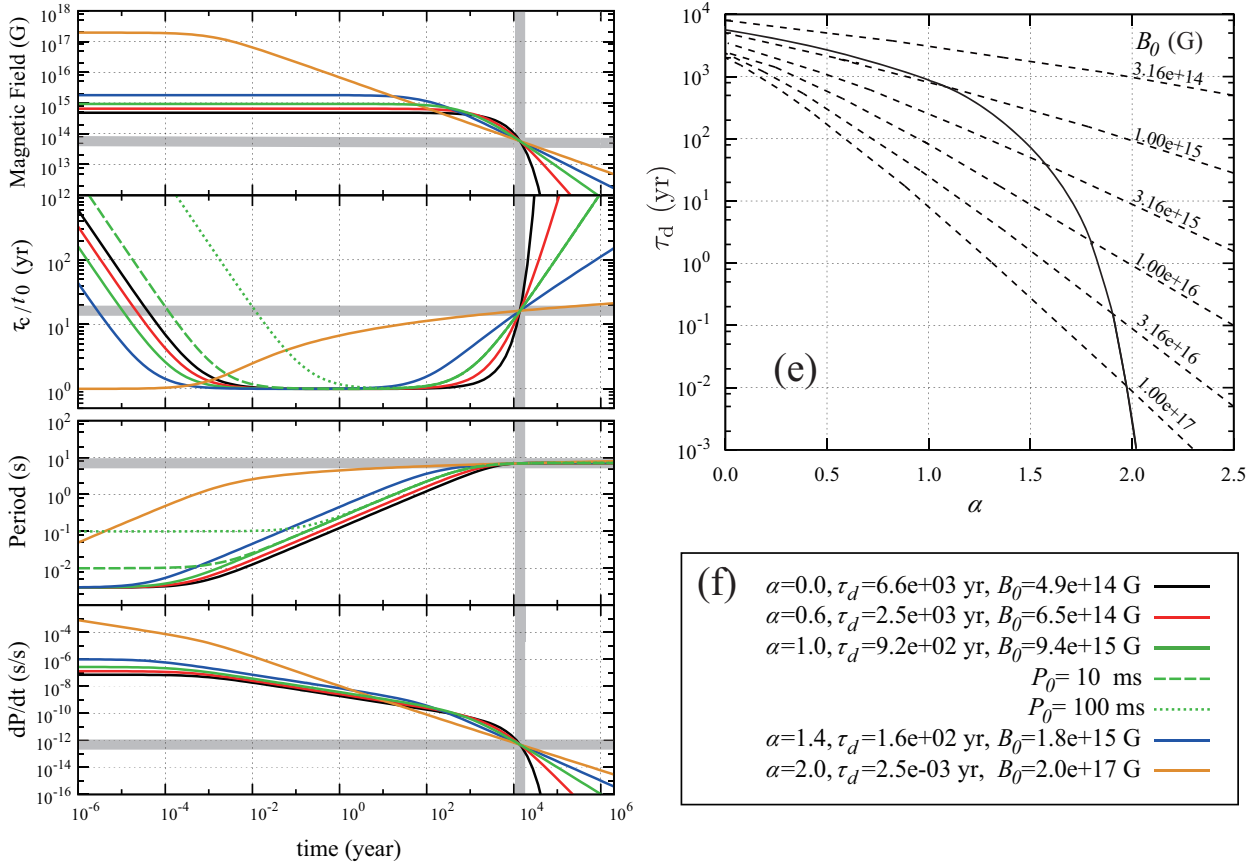} 
	   		\caption{Possible evolution tracks of 1E~2259+586 assuming equation (\ref{eq:domega/dt}) and equation (\ref{eq:dB/dt}). 
				Panels (a)-(d) represent the behavior of the magnetic field $B$, the over-estimation factor of the characteristic age (i.e., $\tau_\mathrm{c}/t_0$),
				the pulse period $P$, and its time derivative $\dot{P}$, respectively.
				The six representative tracks are all constrained to reproduce the presently measured $P$ and $\dot{P}$ at $t = 14$~kyr.
				The dashed and dotted ones assume $P_0=10~\mathrm{ms}$ and $100~\mathrm{ms}$, respectively, while the other five all $P_0=3~\mathrm{ms}$.
				Panel (e) shows the trajectory of solutions that can explain the present-day ($t = 14$~kyr) 1E~2259+586.
				Dashed lines indicate the initial filed value $B_0$.
				Panel~(f) summarize the parameter sets of the trajectories.
			}
	   		\label{fig:SpinMag}
	\end{figure}

	Our next task is to examine whether the observed values of  $P$ and $\dot{P}$ of 1E~2259+586 can be explained
		with the picture presented in section \ref{subsec:decay}.
	Equation (\ref{eq:Tauc}) involves four free parameters, namely $\alpha$, $B_0$ , $\tau_\mathrm{d}$ and $P_0$,
		whereas we have only two observables, $P$ and $\dot{P}$ at $t = t_0 \simeq 14$~kyr.  
	In section 5.1, we showed that the effect of $P_0$ can be neglected, 
		when the current $P$ is long enough.
	Therefore, we chose to fix $P_0$ at 3 ms where strong dynamo works efficiently (e.g., \citealt{Duncan1996}).
	To visualize effects of $P_0$, another solution 
		with $P_0=10, 100~\mathrm{ms}$ and $\alpha = 1.2$ is also shown in figure \ref{fig:SpinMag} [panels (b), (c), and (d)].
	Thus, the effects of $P_0$ are limited to very early ($\ll 1~\mathrm{s}$) stages of the evolution, 
		and its value does not affect our discussion as long as it is much shorter than $\sim 6.7$~s.

	Then, if $\alpha$ is specified, we can find a pair $(B_0,\tau_\mathrm{d})$ that can simultaneously explain $P$ and $\dot{P}$ at present.
	Figure \ref{fig:SpinMag} shows the behavior of such a family of solutions to equation (\ref{eq:Tauc}). 
	Below, we try to constrain the values of $\alpha$ (hence of $B_0$ and $\tau_\mathrm{d}$),
		assuming that $\alpha$ is relatively common among magnetars.
	This is because the broad-band X-ray spectra of magnetars are determined rather uniquely by $\tau_\mathrm{c}$ \citep{Enoto2010b},
		so that $\tau_\mathrm{c}$ is considered to be tightly related to $t_0$ 
		even if these two are unlikely to be identical:  object-to-object scatter in $\alpha$ would cause a scatter in the $\tau_\mathrm{c}/t_0$ ratio, 
		and would make the relation of \citet{Enoto2010b} difficult to interpret.

	When $\alpha$ is small ($0 \leq \alpha < 0.5$), 
		the field would decay, as seen in equation (10),
		either exponentially (if $\alpha=0$) on a time scale of $\tau_{\rm d}$,
		or if $\alpha\ne 0$,
	with a relatively steep power-law after a long lead time $\tau_{\rm d} \sim t_0$.
	The required initial field, $B_0 \sim 10^{15}$ G, is reasonable.
	However, the implied view would be rather ad-hoc:
		1E 2259+586 had been relatively inactive until recently,
		when it suddenly started to  release its magnetic energy with a high rate.
	Furthermore, if such a small value of $\alpha$ were common to magnetars,
		their age differences would make, as in figure 3 (a),
		their present-day field distribution scatter much more widely than is observed.
	Hence we regard these small values of $\alpha$ unlikely.

	As $\alpha$ increases towards 2.0, 
		the power-law field decay becomes milder,
		with shorter values of $\tau_{\rm d}$
		and stronger initial fields $B_0$.
	The implication is that the object started releasing its 
		magnetic energy rather soon after the birth,
		and had already dumped  away a large fraction of its rotational energy
		at a very early stage when the field was still very strong.
	As seen in figure 3 (c),
		the spin period has almost converged to its terminal value (see also \citealt{DallOsso2012}).
	Therefore, this case can explain the observed 
		narrow scatter in $P$ of magnetars,
		assuming that they share relatively similar values of $\alpha$ and $B_0$.
	However,  the cases with  $\alpha \sim 2.0$ or larger would require
		too strong initial fields,
		e.g., $B_0 \sim 10^{17}$ G,
		which would be much higher than the strongest dipole field  
		observed from magnetars,
		$B=2.4 \times 10^{15}$~G of SGR~1806$-$20 (Nakagawa et al. 2009).
	Therefore, such large-$\alpha$ solutions are unlikely, too.

	To summarize these examinations,
		figure 3 (e) shows the locus of the allowed solutions 
		on the $(\alpha, \tau_{\rm d})$ plane, 
		where the values of $B_0$ are also indicated.
	We thus reconfirm the above considerations, 
		that the range of $1\lesssim \alpha < 2$ is appropriate,
		in agreement with the suggestion by \cite{DallOsso2012}.
	Some discussions follow in subsection \ref{subsec:CWOO}.
	
\section{Discussion}
\subsection{Comparison with Other Objects}
\label{subsec:CWOO}
	We reconfirmed the age problem of 1E~2259+586 and CTB~109,  and presented a way to solve it with a simple magnetic field decay model.
	The result agrees with the basic concept of magnetar hypothesis 
		which implies 
			that the energies stored by their magnetic fields should be consumed to supply their X-ray luminosities exceeding those available with their spin down.
	The amount of released field energies can be reflected in the overestimations of the characteristic ages.
	
\begin{figure}[htbp] 
	 	\centering
		\includegraphics[width=4in]{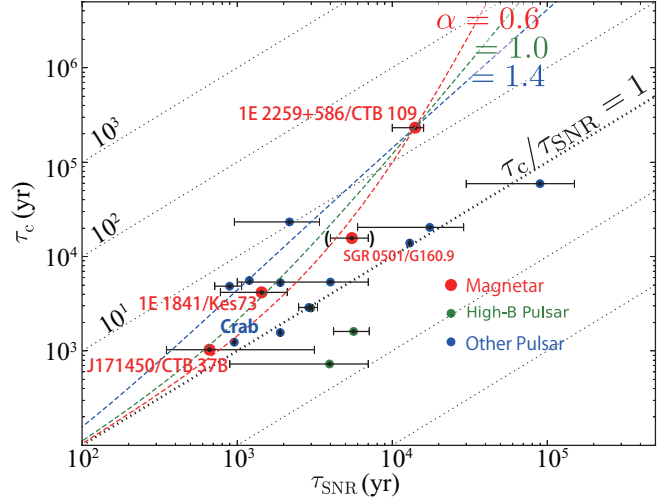} 
		\caption{Relations between $\tau_\mathrm{SNR}$ and $\tau_\mathrm{c}$ of single NSs associated with SNRs. 
			Red, magenta and blue represent magnetars,  high-B pulsars and rotation powered pulsars, respectively.
			Parameters are listed in table \ref{tab:Parameters}.
			The SGR~0501+4516/G160.9+2.6 pair is parenthesized, 
				because the association is rather doubtful, and this SNR might be associated to another pulsar PSR B0458+46 (e.g., \citealt{Leahy1991}). 
			The red, green, and blue dashed curves indicate solutions to equation (\ref{eq:Tauc}),
			with $(\alpha, \tau_\mathrm{d}, B_0) =$ (0.6, $2.5 \times 10^3$~yr, $6.5 \times 10^{14}$~G), (1.0, $9.2\times 10^{2}$~yr, $9.4\times 10^{15}$~G) and (1.4, $1.6\times 10^{2}$~yr, $1.8\times 10^{15}$~G), respectively.
			They all assume $P_0 = 3$~ms, and $B_0$ which is specified by figure \ref{fig:SpinMag} (e).
		}
		\label{fig:SnrAgeVsPsrAge}
	\end{figure}

	Let us then examine whether this concept applies to other NS/SNR associations,
		including both magnetars and ordinary pulsars.
	Figure \ref{fig:SnrAgeVsPsrAge} shows relations between $\tau_\mathrm{c}$ of such single pulsars and the ages of their host SNRs.
	Data points of ordinary pulsars are distributed around the line representing $\tau_\mathrm{c}/\tau_\mathrm{SNR} = 1$ 
		with a few exceptions (e.g., \citealt{Torii1999} for J1811$-$1925/G11.2$-$0.3).
	Therefore, radio pulsars, including the particular case of the Crab pulsar (section \ref{subsec:constB}),
		are considered to be free from the age problem. 

	In addition to the ordinary pulsars, figure \ref{fig:SnrAgeVsPsrAge} shows a few other magnetar/SNR associations.
	The magnetar CXOU J171405.7-381031 has a very small characteristic age of  0.96~kyr \citep{Sato2010}, 
		which is consistent, within rather large errors, with the age ($0.65^{+2.5}_{-0.3}$~kyr; \citealt{Nakamura2009})
		the associated SNR, CTB~37B.
	Another magnetar/SNR association, 1E~1841-045/Kes73, is located in figure \ref{fig:SnrAgeVsPsrAge} between J171405.7-381031/CTB~37B and 1E~2259+586/CTB~109.
	The age of Kes73 was estimated by \cite{kumar2014}, 
		as $0.75-2.1~\mathrm{kyr}$ (table \ref{tab:Parameters}).
	Combining this with $\tau_\mathrm{c} = 4.7~\mathrm{kyr}$ of 1E~1841-045 (table \ref{tab:Parameters}), 
		the age discrepancy of this pair becomes $\tau_\mathrm{c}/\tau_\mathrm{SNR} = 2.7-8$.
	These two associations do not show large overestimation factors of $\tau_\mathrm{c}/\tau_\mathrm{SNR}$ as much as 1E~2259+586/CTB~109 association.
	Thus, the three magnetar/SNR associations (including 1E~2259+586/CTB~109) suggest that the age over-estimation factor,
		 $\tau_\mathrm{c}/\tau_\mathrm{SNR}$, 
		 increases towards older objects.
	This agrees, at least qualitatively, with the theoretical behavior seen in figure \ref{fig:SpinMag} (c), as long as $P_0 $ is negligible.
	
	We hence tried to explain the data points of these three magnetar/SNR associations 
		with a common set of parameters, 
		and derive a plausible range of $\alpha$.
	For this purpose, three evolution tracks representing the solutions to equation  (\ref{eq:Tauc}) for 1E~12259+586/CTB~109 are additionally plotted on figure \ref{fig:SnrAgeVsPsrAge}.
	Each parameter set is the  same as that of figure \ref{fig:SpinMag}(f).
	If $\alpha$ is small as $0 \leq \alpha < 0.6$ (dashed red line), the CXOU~J171405.7-381031/CTB~37B association cannot be explained.
	On the other hand, large value of $\alpha (>1.5)$ fail to explain the 1E~1841-045/Kes73 association.
	Thus, the three magnetar/SNR pairs in figure \ref{fig:SnrAgeVsPsrAge} can be explained in a unified way if they have a common value of $\alpha$ (and also of $\tau_\mathrm{d}$)
		that is in the range of $0.6 < \alpha < 1.4$.

\subsection{Supporting Evidence}
	
\begin{figure}[htbp] 
		\centering
		\includegraphics[width=5in]{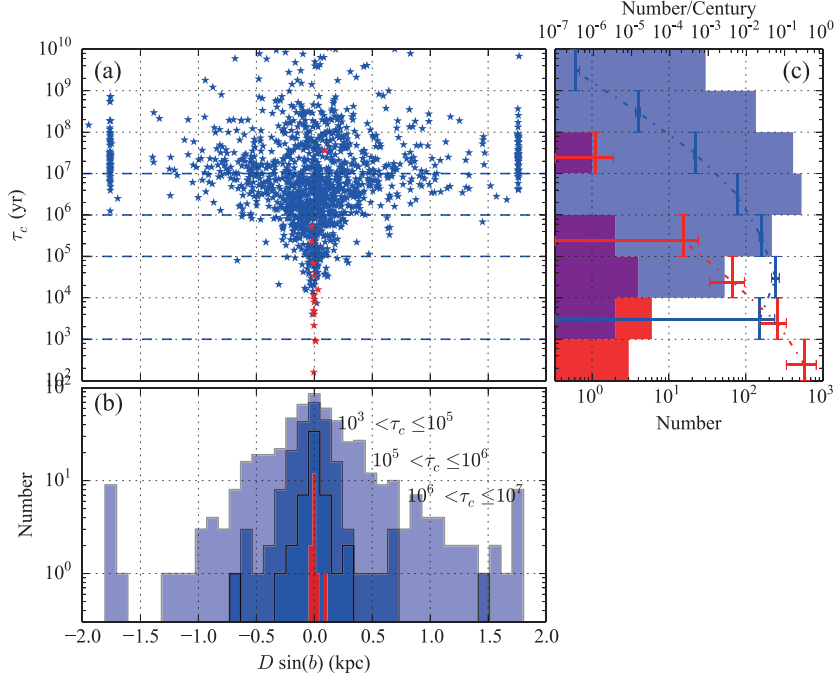} 
		\caption{ (a) Spatial distributions of magnetars (red) and radio pulsars (blue).
		 	Abscissa and ordinate means distance from the Galactic plane and characteristic age, respectively. 
			(b) Projection of panel (a) onto the direction perpendicular to the Galactic plane.
			Radio pulsars are divided into three subgroups according to their age.
			(c) Age distributions of the objects, produced by projecting panel (a) onto the time axis.
			Histograms represent numbers of pulsars with ages in that logarithmic interval,
			while crosses tied by a dotted line show the object number per century.
		}
		\label{fig:GalacticDistribution}
	\end{figure}

	The scenario so far developed implies that magnetars form a population 
		that is much younger than previously thought .
	This important inference is supported by an independent piece of evidence.
	Figure \ref{fig:GalacticDistribution} shows a spatial distribution of NSs including magnetars. 
	Because of steady motions after kick velocities are given by explosions, older pulsars with larger $\tau_\mathrm{c}$ are thus distributed up to farther distances from the Galactic plane.
	In contrast, magnetars are much more concentrated to the plane for their nominal age,
		as better seen in figure \ref{fig:GalacticDistribution} (b) which is a projection of figure \ref{fig:GalacticDistribution} (a) along the direction perpendicular to the Galactic plane.
	This implies two possible scenarios; 
		magnetars, as we have shown, are much younger than indicated by their $\tau_\mathrm{c}$,
		or their kick velocities are systematically lower than those of others.  
	Recently, proper motions of four magnetars (SGR~1806$-$20, SGR~1900+14, 
		1E~2259+586 and 4U~0142+61) were successfully measured by \cite{Tendulkar2012, Tendulkar2013}.
	They calculated the mean and standard deviation of their ejection velocities as $200~\mathrm{km~s}^{-1}$ and $90~\mathrm{km~s}^{-1}$, respectively.
	They also conclude 
		that the weighted average velocity of magnetars is in good agreement 
		with the tangential velocities of the pulsar population \citep{Hobbs2005}.
	Therefore, we are left with the former of the two possibilities. 
	In other words,  magnetars should be  systematically younger than ordinary pulsars that have similar $\tau_\mathrm{c}$.


\subsection{Implication for the Magnetar Population}	
	Observationally, magnetars are no longer a minority of NS species.  This is shown in figure \ref{fig:GalacticDistribution} (c), 
		which is the projection of figure \ref{fig:GalacticDistribution} (a) onto the time axis.
	Thus, even if $\tau_\mathrm{c}$ is not corrected for the over estimation, 
		magnetars already occupy a considerable fraction of young NSs.
	If we replace $\tau_\mathrm{c}$ of magnetars with their true ages, 
		their dominance among young NSs will become even more enhanced.

	As yet another important implication, we expect that numerous aged magnetar descendants would lurk in our Galaxy.
	Radio pulsars, observed as the major population of NSs, cannot harbor such aged magnetars, because their $P$ are shorter than those of magnetars.
	Instead, such objects may be being discovered as weak-field SGRs, including SGR~0418+5729 \citep{Rea2013},
	Swift~J1822.3−1606 (\citealt{Rea2012}; \citealt{Scholz2012}) and 3XMM J185246.6+003317 \citep{Zhou2014}.

\section{Summary}

	We performed 4 pointings observations of CTB~109 with Suzaku.
	The spectra extracted from eastern parts of the SNR were well fitted with two plasma components having two different temperatures.
	Assuming thermal equilibrium between electrons and protons, the shock velocity was calculated as $460~{\rm km~s^{-1}}$,
	 and the age of the SNR was estimated as $14~{\rm kyr}$ using the Sedov-similarity solution.
	These results are consistent with the conclusion of the previous work by \cite{Sasaki2013}.
	We thus reconfirmed the huge discrepancy between the age of CTB~109 and the characteristic age of 1E~2259+586.

	We consider that the characteristic age of 1E 2259+586 is significantly overestimated,
		as compared to its true age which we identify with that of CTB109.
	This effect, seen also in some other magnetars to a lesser extent, 
		can be attributed to decay of their magnetic fields,
		as implied by the basic concept of ``magnetars".
	In fact, 
		the observed pulse period and its derivative of 1E 2259+586 has been explained successfully
		by a family of solutions to a simple equation describing the magnetic field decay.
	Furthermore,
		the $\tau_\mathrm{c}$ vs $\tau_\mathrm{SNR}$ relations of the three magnetar-SNR associations, 
		including the 1E~2259+586/CTB~109 pair,
		can be explained consistently if they have a common value of $\alpha$ in the range of 0.6-1.4. 
	As a result, 
	magnetars are considered to be much younger than was considered so far,
		and are rather dominant among new-born NSs.
	The youth of magnetars is supported independently by their much stronger concentration along the Galactic plane than  ordinary pulsars.

	
\newpage
\appendix
\begin{table}
	\begin{center}
		\caption{Parameters for Figure \ref{fig:SnrAgeVsPsrAge}. }
		\begin{tabular}{lcccccc}
   			\hline
			\multicolumn{1}{c}{\#}       &Pulsar/SNR  &  P~(ms) & $\dot{P}$~(ss$^{-1}$)    & B~($\times10^{12}$ G)  &  $\tau_\mathrm{c}$~(kyr)   &  $\tau_\mathrm{SNR}$~(kyr)\\
  			\hline
1 &	1E 1841-045/Kes73 &	11778 &	$4.5\times10^{-11}$ 		&	730 &	4 &	0.75 - 2.1 \\
2 &	SGR~0501+4516/G160.9+2.6&	5762 &	$5.8\times10^{-12}$ &	190 &	16 &	4 - 7 \\
3 &	J171405.7-381031/CTB37B &	3824 &	$5.9\times10^{-11}$ &	480 &	0.96 &	$0.65^{+25}_{-0.3}$ \\
4 &	1E 2259+586/CTB~109\footnotemark[$*$] &	6979 &	$4.8\times10^{-13}$ &	58 &	230 &	10 - 16 \\
\\
5 &	J1846-0258/Kes75 &	326 &	$7.1\times10^{-12}$ 		&	49 &	0.7 & 0.9 - 4.3 \\
6 &	J1119-6127/G292.2-0.5 &	407 &	$4.0\times10^{-12}$ &	41 &	1.6 &	 4.2 - 7.1 \\
7 &	J1124-5916/G292.0+1.8 &	135 &	$7.5\times10^{-13}$ &	10 &	2.9 &	2.93 - 3.05 \\
\\
8 &	J1513-5908/G320.4-1.2 &	151 &	$1.5\times10^{-12}$ &	15 &	1.6 &	1.9 \\
9 &	J0007+7303/G119.5+10.2 &	315 &	$3.6\times10^{-13}$ &	11 &	14 &	13 \\
10 &	J1930+1852/G54.1+0.3 &	136 &	$7.5\times10^{-13}$ &	10 &	2.9 &	2.5 - 3.3 \\
11 &	J1856+0113/W44 &	267 &	$2.1\times10^{-13}$ 		&	7.5 &	20 &	6 - 29 \\
12 &	J0633+0632/G205.5+0.5 &	297 &$8.0\times10^{-14}$&	4.9 &	60 &	30 - 150 \\
13 &	Crab &	33 &				$4.2\times10^{-13}$ 		&	3.8 &	1.2 &	0.959 \\
14 &	J0205+6449/3C 58 &	65 &	$1.9\times10^{-13}$ 		&	3.6 &	5 &	1-7 \\
15 &	J1833-1034/G21.5-0.9 &	61 &	$2.0\times10^{-13}$ 		&	3.6 &	5 &	0.72 - 1.07 \\
16 &	J1747-2809/G0.9+0.1 &	52 &	$1.6\times10^{-13}$ 		&	2.9 &	5 &	1.9 \\
17 &	J1813-1749/G12.8-0.0 &	44 &	$1.3\times10^{-13}$ 		&	2.4 &	6 &	1.2 \\
18 &	J1811-1925/G11.2-0.3 &	64 &	$4.4\times10^{-14}$ 		&	1.7 &	2.3 & 0.96 - 3.4 \\		\hline
			\multicolumn{4}{@{}l@{}}{\hbox to 0pt{\parbox{170mm}{
				\footnotesize
				\par\noindent
				\footnotemark[$*$] This work.\\
				Data for $P$ and $\dot{P}$ of pulsar were collected from ATNF Pulsar catalogue \citep{Manchester2005}\footnotemark[1] \\
				Data for $\tau_\mathrm{SNR}$ were collected from  \cite{Ferrand2012} \footnotemark[2]
			}\hss}}
		\end{tabular}
		\label{tab:Parameters}
 	\end{center}
 \end{table}
 \footnotetext[1]{http://www.atnf.csiro.au/research/pulsar/psrcat}
 \footnotetext[2]{http://www.physics.umanitoba.ca/snr/SNRcat} 




\def\aj{AJ}%
\def\actaa{Acta Astron.}%
\def\araa{ARA\&A}%
\def\apj{ApJ}%
\def\apjl{ApJ}%
\def\apjs{ApJS}%
\def\ao{Appl.‾Opt.}%
\def\apss{Ap\&SS}%
\def\aap{A\&A}%
\def\aapr{A\&A‾Rev.}%
\def\aaps{A\&AS}%
\def\azh{AZh}%
\def\baas{BAAS}%
\def\bac{Bull. astr. Inst. Czechosl.}%
\def\caa{Chinese Astron. Astrophys.}%
\def\cjaa{Chinese J. Astron. Astrophys.}%
\def\icarus{Icarus}%
\def\jcap{J. Cosmology Astropart. Phys.}%
\def\jrasc{JRASC}%
\def\mnras{MNRAS}%
\def\memras{MmRAS}%
\def\na{New A}%
\def\nar{New A Rev.}%
\def\pasa{PASA}%
\def\pra{Phys.‾Rev.‾A}%
\def\prb{Phys.‾Rev.‾B}%
\def\prc{Phys.‾Rev.‾C}%
\def\prd{Phys.‾Rev.‾D}%
\def\pre{Phys.‾Rev.‾E}%
\def\prl{Phys.‾Rev.‾Lett.}%
\def\pasp{PASP}%
\def\pasj{PASJ}%
\def\qjras{QJRAS}%
\def\rmxaa{Rev. Mexicana Astron. Astrofis.}%
\def\skytel{S\&T}%
\def\solphys{Sol.‾Phys.}%
\def\sovast{Soviet‾Ast.}%
\def\ssr{Space‾Sci.‾Rev.}%
\def\zap{ZAp}%
\def\nat{Nature}%
\def\iaucirc{IAU‾Circ.}%
\def\aplett{Astrophys.‾Lett.}%
\def\apspr{Astrophys.‾Space‾Phys.‾Res.}%
\def\bain{Bull.‾Astron.‾Inst.‾Netherlands}%
\def\fcp{Fund.‾Cosmic‾Phys.}%
\def\gca{Geochim.‾Cosmochim.‾Acta}%
\def\grl{Geophys.‾Res.‾Lett.}%
\def\jcp{J.‾Chem.‾Phys.}%
\def\jgr{J.‾Geophys.‾Res.}%
\def\jqsrt{J.‾Quant.‾Spec.‾Radiat.‾Transf.}%
\def\memsai{Mem.‾Soc.‾Astron.‾Italiana}%
\def\nphysa{Nucl.‾Phys.‾A}%
\def\physrep{Phys.‾Rep.}%
\def\physscr{Phys.‾Scr}%
\def\planss{Planet.‾Space‾Sci.}%
\def\procspie{Proc.‾SPIE}%


\end{document}